\documentclass[10pt]{iopart}

\usepackage{graphicx}
\usepackage{hyperref}
\usepackage{amssymb}
\usepackage{longtable}
\usepackage{rotating}
\usepackage{color}
\usepackage{epsfig}
\usepackage{epsf}
\usepackage{fancyhdr}

\def\be{\begin{equation}}
\def\ee{\end{equation}}
\def\bea{\begin{eqnarray}}
\def\eea{\end{eqnarray}}

\def\sci#1#2{#1\times10^{#2}}

\def\th{\textrm{\mbox{\tiny{th}}}}

\def\Tcoh{T_{\textrm{\mbox{\tiny{coh}}}}}

\def\th{\textrm{\mbox{\tiny{th}}}}

\newcommand{\F}{\mathcal{F}}
\begin{document}
%
\title[A $\chi^2$ veto for continuous gravitational wave searches]
{A $\chi^2$ veto for continuous gravitational wave searches}

\author{Llucia Sancho de la Jordana\dag\ and Alicia M. Sintes\dag\ddag}

\address{\dag\ Departament de F\'{\i}sica, Universitat de les Illes
Balears, Cra. Valldemossa Km. 7.5, E-07122 Palma de Mallorca,
Spain}
\address{\ddag\ Max-Planck-Institut f\"ur
    Gravitationsphysik, Albert Einstein Institut, Am M\"uhlenberg 1,
    D-14476 Golm, Germany}

\address{E-mail: \texttt{llucia.sancho@uib.es}, \texttt{sintes@aei.mpg.de} }

\begin{abstract}

$\chi^2$ vetoes are commonly used in searching for gravitational waves, 
in particular for broad-band signals, but they can also be applied to
narrow-band continuous wave signals, such as those expected from rapidly 
rotating neutron stars. 
In this paper we present a $\chi^2$ veto adapted to the Hough transform searches for continuous gravitational wave signals;
we characterize the $\chi^2$-significance plane for different frequency bands; and discuss the expected performance 
of this veto in  LIGO analysis.
\end{abstract}

\date{\today}

\pacs{04.80.Nn, 07.05.Kf, 95.55.Ym, 97.60.Gb}
%

\section{Introduction}
\label{sec:intro}

Continuous gravitational wave signals emitted by neutron stars in our galaxy
are among the targets of on-going searches for gravitational waves using GEO600 \cite{GEO1,GEO2}, 
LIGO \cite{ligo1,ligo2} and VIRGO \cite{virgo97} data.
Examples of such searches include targeting known radio pulsars \cite{S1PulsarPaper,S2TDPaper,S3S4TDPaper}
and the low-mass X-ray binary system Scorpius X-1 \cite{S2FstatPaper, S4RadiometerPaper}, as well as  all-sky surveys for unknown rotating neutron stars \cite{S2FstatPaper,S2HoughPaper,S4IncoherentPaper}. 
The first type of searches typically use matched filtering techniques and are not very computationally expensive. The second type of searches look for as yet undiscovered sources. This involves searching over large parameter space volumes and turns out to be computationally limited, as the number of templates  that must be searched over increase rapidly with the observation time.
The ultimate goal for wide parameter searches for continuous signals over large data sets is to employ hierarchical schemes  which alternate coherent and semi-coherent techniques \cite{bc, pss01,cgk,f2000,fap,fp}, as those currently employed by Einstein@Home \cite{EatH}, a distributed-computing effort that uses the idle CPU time of computers across the world.

The Hough transform \cite{hough04,ks07}
is an example of a semi-coherent method that can be used to select candidates in parameter space to be followed up. 
Results of the Hough transform to search the entire sky have been reported in
 \cite{S2HoughPaper,S4IncoherentPaper}. In those papers, the Hough transform was used to search for cumulative excess power from a hypothetical periodic signal by examining successive spectral estimates based on short Fourier transforms (SFTs)
of the calibrated detector data, taking into account the Doppler frequency shift due to the motion of the detector  with respect to the solar system barycenter and the intrinsic frequency evolution of the source.
Two flavors of the Hough transform have been developed and employed for different searches, the 'standard Hough' \cite{hough04,S2HoughPaper} and the 'weighted Hough' \cite{ks07,S4IncoherentPaper}. In the 'standard Hough',
the cumulative excess power is computed as 
 the sum of binary zeroes and ones, where a SFT contributes unity if the power exceeds a normalized threshold, and in the 'weighted Hough' the contribution of the SFTs is weighted according to the noise and detector antenna pattern to maximize the signal-to-noise (SNR) ratio.

In all hierarchical methods it is crucial that the selection of candidates done by the semi-coherent stage is as effective as possible, since it determines the final sensitivity of the full pipeline.
For this reason the development of veto and/or coincidence tests is very important  in order to reduce the number of false alarms. In this paper we present a $\chi^2$ veto adapted to the Hough transform searches for continuous wave signals  and we discuss the expected performance of this veto in  LIGO analysis.

$\chi^2$ discriminators are commonly used in searching for gravitational wave
signals.  In particular, for binary inspiral searches  a
$\chi^2$ time-frequency discriminator is used as a veto for the output of matched
filter, by analyzing the output of different frequency bands. This $\chi^2$ test
\cite{grasp,allen05} was specifically constructed for broadband signals, but it can be
modified for signals that are narrow band as the
continuous wave signals expected from rapidly rotating 
neutron stars.
For these continuous wave signals that will be observed over periods of several months, 
in order to build up the sufficient SNR,
we propose to split the data into several chunks, analyze each chunk separately 
and construct a $\chi^2$ statistic by combining the partial results.
This statistic would then be able to discriminate if the SNR accumulates 
along the different chunks in a way that is consistent with the properties of the signal
and the detector noise. 

The rest of the paper is organized as follows: Section \ref{sec:hough}  briefly summarizes the Hough transform and  its statistical properties. Section \ref{sec:chi} derives a $\chi^2$ discriminator for different implementations of the Hough transform. Section \ref{sec:s4} characterizes the $\chi^2$-significance plane and discusses its application using LIGO data.
Section \ref{sec:conc} concludes with a summary of the results.


\section{The Hough transform method}
\label{sec:hough}

The Hough transform is a well known method for pattern recognition  that has been applied to the search for continuous gravitational waves.
In this case the Hough transform is used to find a signal whose frequency evolution fits the pattern produced by the Doppler shift and the spin-down in the time-frequency plane of the data. Further details can be found in  \cite{hough04}, here we only give a brief summary  and statistical properties.

The starting point for the Hough transform are $N$ SFTs.
Each of these SFTs is digitized by setting a threshold $\rho_\th$ on the normalized power
\begin{equation} \label{eq:normpower}
\rho_k = \frac{2|\tilde{x}_k|^2}{\Tcoh S_n(f_k)} \,.
\end{equation}
Here  ${\tilde{x}_k}$ is the discrete Fourier transform of the data,
the frequency index $k$ corresponds to a physical frequency of $f_k= k/\Tcoh$,
$S_n(f_k)$ is the single sided power spectral density of the
detector noise and $\Tcoh$ is the time baseline of the SFT.
The $k^{th}$ frequency bin is selected if $\rho_k
\geq \rho_\th$, and rejected otherwise.  In this way, each SFT is replaced
by a collection of zeros and ones called a peak-gram. The probability that a frequency bin is selected is  $q = e^{-\rho_\th}$
for Gaussian noise and  $\eta$, given by
\begin{equation}
\label{eq:eta}
\eta = q\left\{1+\frac{\rho_\th}{2}\lambda_k +
\mathcal{O}(\lambda_k^2)  \right\}
\end{equation}
in the presence of a signal. $\lambda_k$ is the signal to noise ratio within a single SFT, and
for the case when there is no mismatch between the signal and the
template:
\begin{equation}
\label{eq:lambda}
\lambda_k = \frac{4|\tilde{h}(f_k)|^2}{\Tcoh S_n(f_k)}
\end{equation}
with $\tilde{h}(f)$ being the Fourier transform of the signal $h(t)$.

The Hough transform is used to map points from the time-frequency plane of our data
(understood as a sequence of peak-grams) into the space of the source parameters.
 Each point in parameter space corresponds to a pattern in the time-frequency plane, 
and  the Hough number count $n$  is  the weighted sum of the ones and zeros
of the different peak-grams along this curve. For the 'weighted Hough' this sum is computed as
\be \label{eq:sumw}
n=\sum_{i=1}^N{w_i n_i}     \, , 
\ee
where $n_i$ is  either 0 or 1 depending on where the power crosses the
threshold and the weights are normalized according to
\be
\sum_{i=1}^N{w_i}  =N \, .
\ee
When  all $w_i=1$  we obtain the 'standard Hough'.
The 'weighted Hough' can improve the sensitivity of the search taking into account
the possible non-stationarities of the detector noise and the amplitude modulation due to the motion of the detector,
 and allow for multi-interferometer searches.

For large values of $N$, the number count distribution $n$ can be considered a continuous
variable and well approximated by a Gaussian distribution:
\be
p(n)=\frac{1}{\sqrt{2 \pi \sigma^2}} \exp \left( - \frac{(n-\langle n\rangle)^2}{2\sigma^2}\right) 
\, .
\ee 
In the absence of a signal, the mean and variance are
\be
\langle n\rangle = Nq \quad \textrm{and} \quad 
\sigma^2= q(1-q)  \sum_{i=1}^Nw_i^2\, ,
\ee
and in the presence of a signal, the mean and variance of $n$ become 
\begin{equation}
  \langle n\rangle = qN + \frac{q\rho_\th}{2}\sum_{i=1}^N w_i\lambda_i
  \qquad \textrm{and} \qquad \sigma^2 =
  \sum_{i=1}^Nw_i^2\eta_i(1-\eta_i)\,. 
\end{equation}
Because of the weights $\sigma$ varies for different sky locations we should not compare number counts directly but the significance of a number count. 
 The significance $s$ of the observed number-counts $n$ is defined as
 \be
 s= \frac{n-\langle n\rangle}{\sigma} \, ,
 \ee
 where $\langle n\rangle$ and $\sigma$ are the expected mean and standard deviation
 for pure noise. Furthermore, one can see that
 setting a threshold at a given false alarm rate $\alpha$ is 
 equivalent to set a threshold at a certain significance \cite{ks07}
 \be
 s_\th= \sqrt{2}\textrm{erfc}^{-1}(2\alpha)\,.
 \ee


\section{The $\chi^2$  veto}
\label{sec:chi}
In this section we derive a  $\chi^2$  discriminator for the different 
implementations of the Hough transform.
The idea is to split the data into $p$  non-overlapping chunks,
each of them containing a certain number of SFTs $\{N_1, N_2,\ldots,N_p\}$, such that 
\be
\sum_{j=1}^p N_j =N \, ,
\ee
and analyze them separately, obtaining the  Hough
number-count $n_j$ which, for the same pattern across the different chunks,
 would then satisfy
\be
\sum_{j=1}^p n_j =n \, ,
\ee
where $n$ is the total number-count for a given point in parameter space.  
The  $\chi^2$ statistic will look along the different chunks to see if the SNR accumulates 
in a way that is consistent with the properties of the signal and 
the detector noise. Small values of $\chi^2$ are consistent with the hypothesis
that  the observed SNR (or more precisely the significance)  arose from a detector output which was
a linear combination of Gaussian noise and the continuous wave signal. 
Large values of $\chi^2$ indicate either the signal did not match the template
or that the detector noise was non-Gaussian.

\subsection{The standard Hough}
\label{sec:chi2SH}
In the simplest case in which all weights are set to unity, it  is easy to see that the expected values of
the number counts are
\be
 \langle n\rangle =N\eta \, ,
 \qquad  
 \sigma^2_n =N\eta(1-\eta) \, ,
 \ee
\be
 \langle n_j\rangle= N_j\eta= N_j\frac{\langle n\rangle}{N} \, ,
 \qquad
 \sigma^2_{n_j} =N_j\eta(1-\eta) \, ,
\ee
where $n$ is the total measured number count and $\eta$ is the probability of selecting
a peak in the presence of a signal.

Consider the $p$ quantities defined by
\be
\Delta n_j\equiv n_j-\frac{N_j}{N}n \, .
\ee
With this definition, it holds true that 
\be\label{eq:chi=0}
\langle \Delta n_j\rangle=0 \, ,
\qquad 
\sum_{j=1}^p \Delta n_j=0 \qquad \,,
\langle  n_j n\rangle= \frac{N_j}{N} \langle n^2\rangle \,,
\ee
and the expectation value of the square of $\Delta n_j$ is
\be
 \langle (\Delta n_j)^2\rangle =\left( 1-\frac{N_j}{N}\right) N_j\eta(1-\eta) \, .
\ee
 Therefore we can define the $\chi^2$ discriminator statistic by
 \be
 \label{eq:chi3}
\chi^2(n_1, \ldots, n_p) =  
\sum_{j=1}^p {\frac{(\Delta n_j)^2}{\sigma^2_{n_j}} }=
\sum_{j=1}^p {\frac{\left(n_j-nN_j/N\right)^2}{N_j\eta(1-\eta)} }\, ,
 \ee
and expected value of $\chi^2$  is
 \be
 \langle \chi^2\rangle=\sum_{j=1}^p \left( 1-\frac{N_j}{N}\right)= p-1 \, .
 \ee
Equation (\ref{eq:chi3}) has a 
$\chi^2$-distribution with $p-1$ degrees of freedom. To implement this discriminator, we need to 
measure, for each point in parameter space, the total number-count $n$, the
partial number-counts $n_j$ and  assume a constant value of $\eta=n/N$.

\subsection{The weighted Hough}
\label{sec:chi2WH}
We can generalize the previous result (\ref{eq:chi3}) 
for the weighted Hough transform. Let $I_j$ be the set of SFT indices for each
different $p$, the mean and variance of the number-count become
\be
\langle  n_j \rangle= \sum_{i\in I_j} w_i\eta_i
\qquad
\langle  n \rangle=   \sum_{j=1}^p \langle  n_j \rangle
\qquad 
\sigma^2_{n_j} = \sum_{i\in I_j} w_i^2\eta_i(1-\eta_i) 
\ee
and we can define
\be
\Delta n_j\equiv n_j-n\frac{\sum_{i\in I_j} w_i\eta_i}{\sum_{i=1}^N w_i\eta_i} \, ,
\ee
 so that 
$\langle \Delta n_j\rangle=0$,
 $\sum_{j=1}^p \Delta n_j=0$. The $\chi^2$ discriminator would now be:
 \be
\chi^2 =  
\sum_{j=1}^p {\frac{(\Delta n_j)^2}{\sigma^2_{n_j}} } =
\sum_{j=1}^p {\frac{\left(n_j-n (\sum_{i\in I_j} w_i\eta_i)/(\sum_{i=1}^N w_i\eta_i)\right)^2}
{\sum_{i\in I_j} w_i^2\eta_i(1-\eta_i)} }\, . \label{eq:chi5}
 \ee
In a given search, we can compute the
 $\sum_{i\in I_j} w_i$, $\sum_{i\in I_j} w_i^2$ for each of the $p$ chunks. 
The problem for implementing this discriminator (\ref{eq:chi5})
  is that the different $\eta_i$ values
 can not be measured from the data itself because they depend on the exact SNR for a single SFT
 as defined in Eqns. (\ref{eq:eta}) and (\ref{eq:lambda}) and not just its averaged value. For this reason 
we propose to approximate  in equation (\ref{eq:chi5}) $\eta_i\rightarrow\eta^*$, where $\eta^*=n/N$. Thus 
 \be
 \label{eq:chi6}
\chi^2 \approx
\sum_{j=1}^p {\frac{\left(n_j-n (\sum_{i\in I_j} w_i)/N\right)^2}
{\eta^*(1-\eta^*) \sum_{i\in I_j} w_i^2 } }\, .
 \ee


\section{Application of the  $\chi^2$ veto on the LIGO fourth science run}
\label{sec:s4}

\begin{figure}
\begin{center}
\begin{tabular}{cc}
\includegraphics[width=6.3cm]{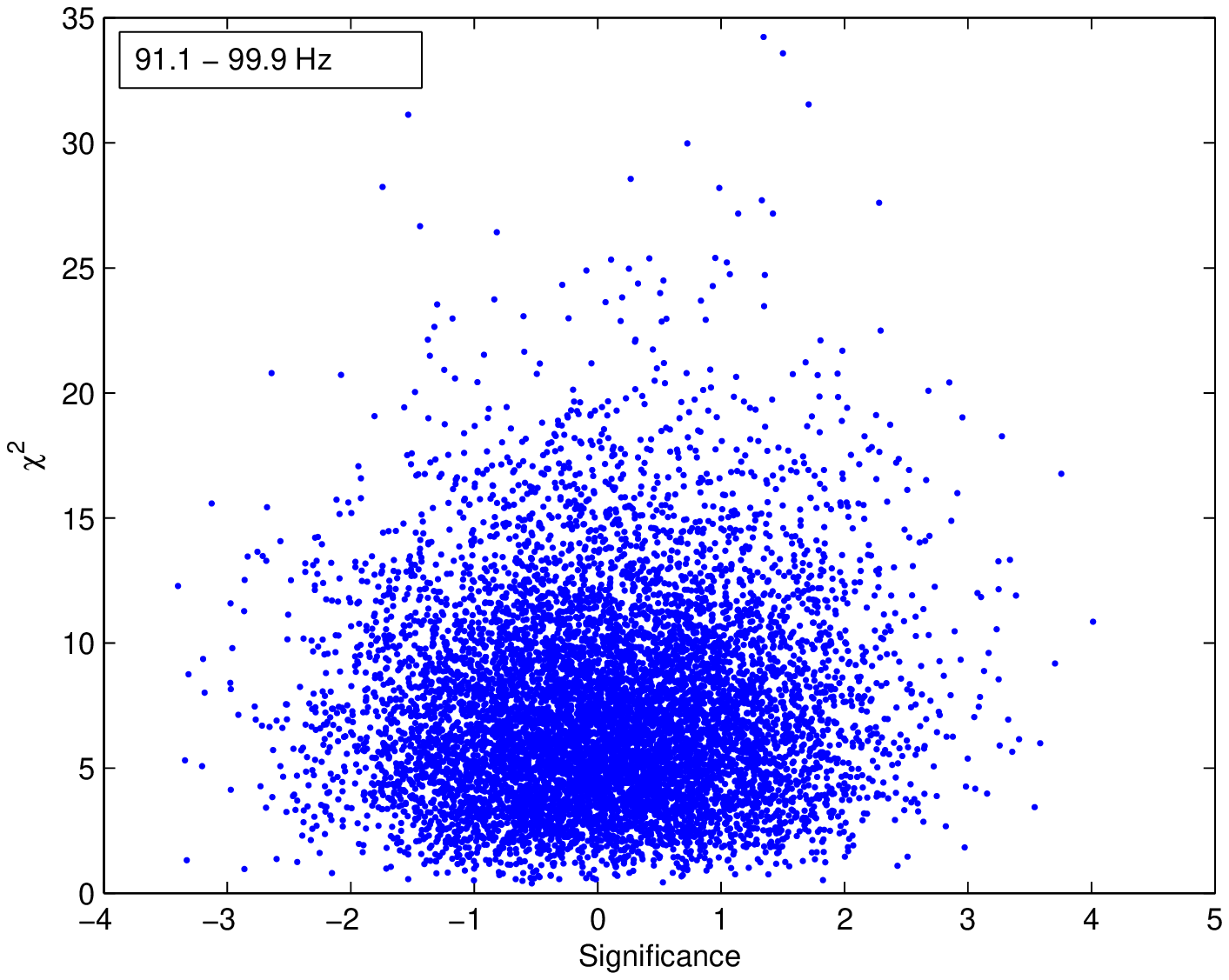} &
\includegraphics[width=6.3cm]{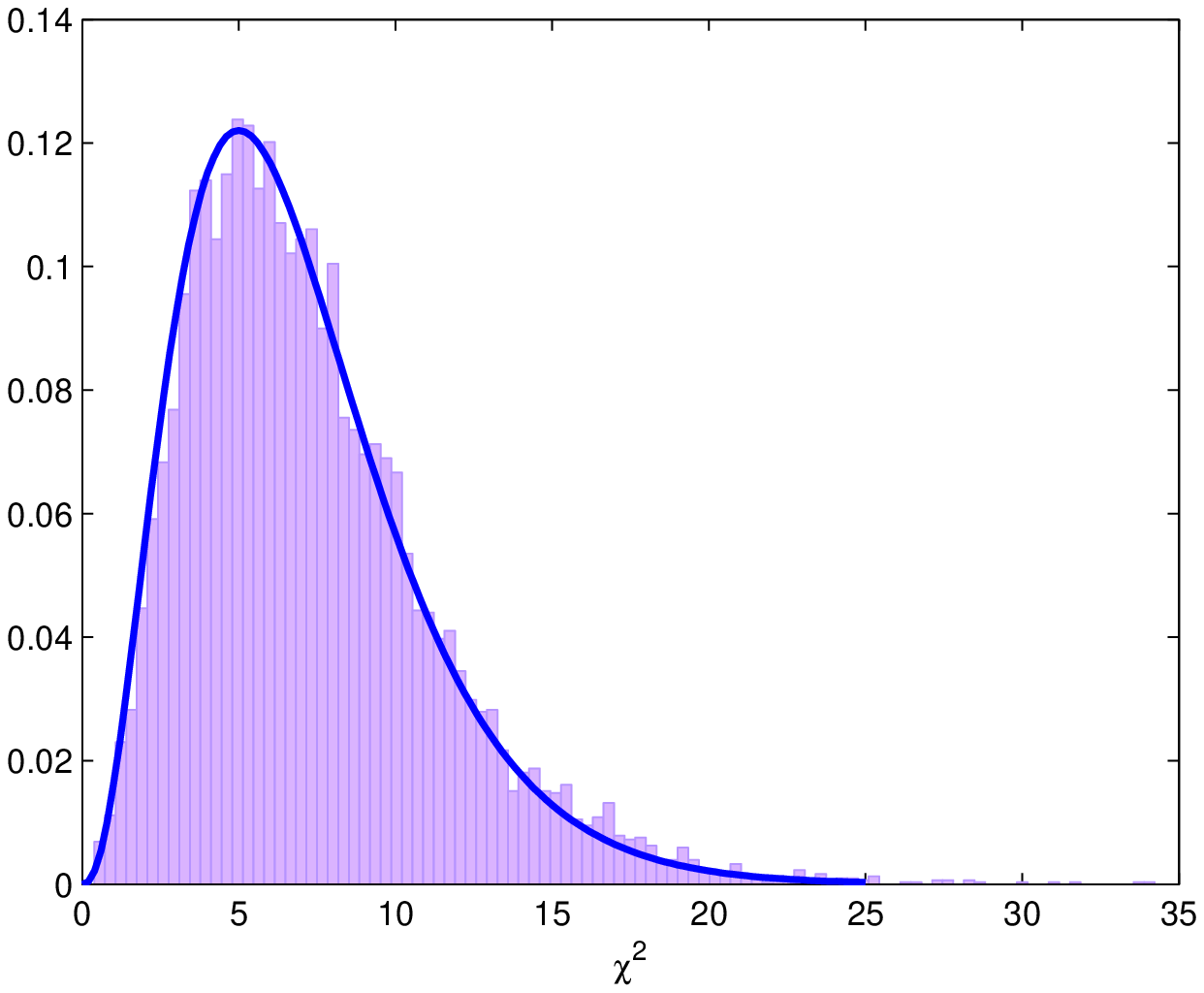} \\
\includegraphics[width=6.3cm]{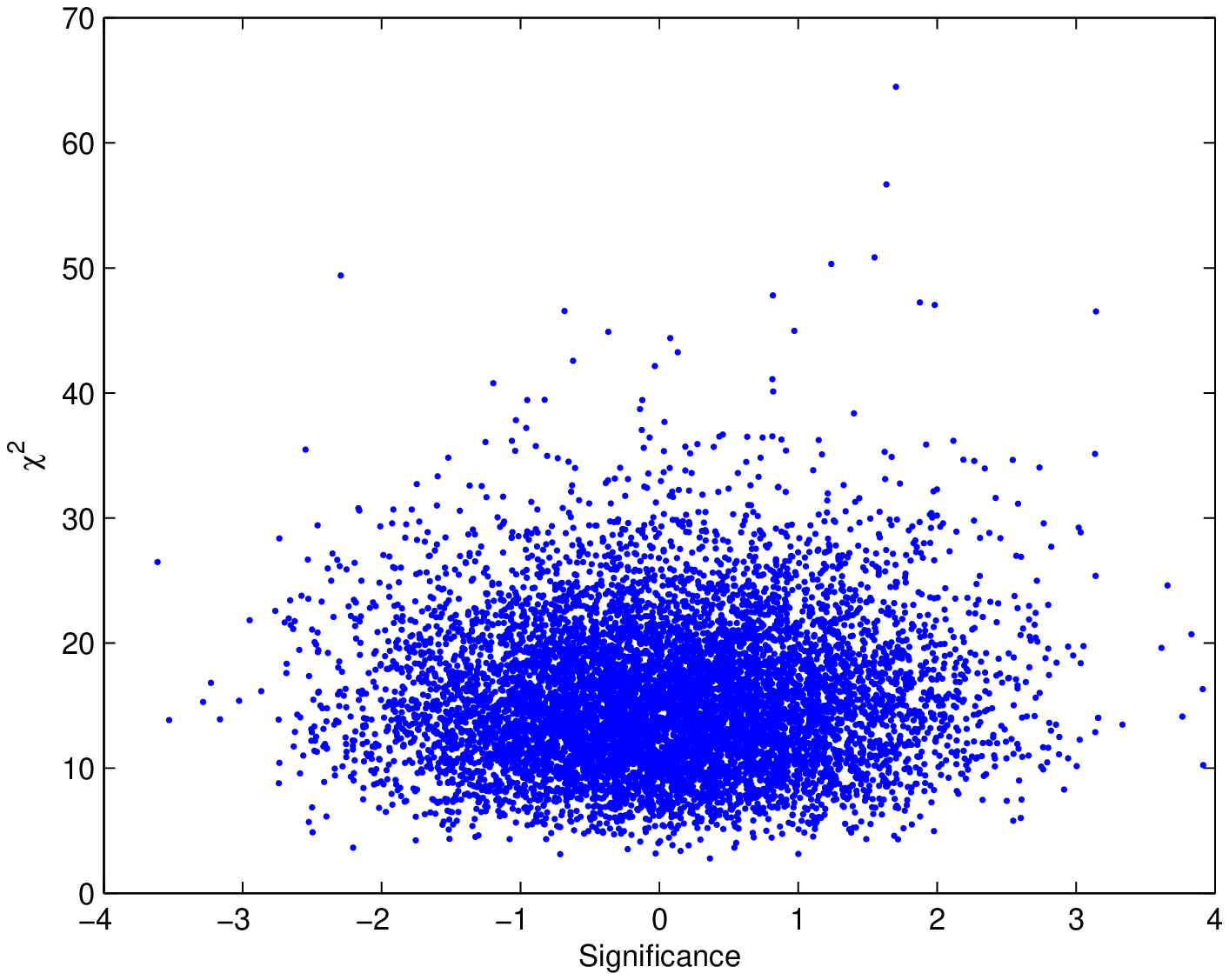} & 
\includegraphics[width=6.3cm]{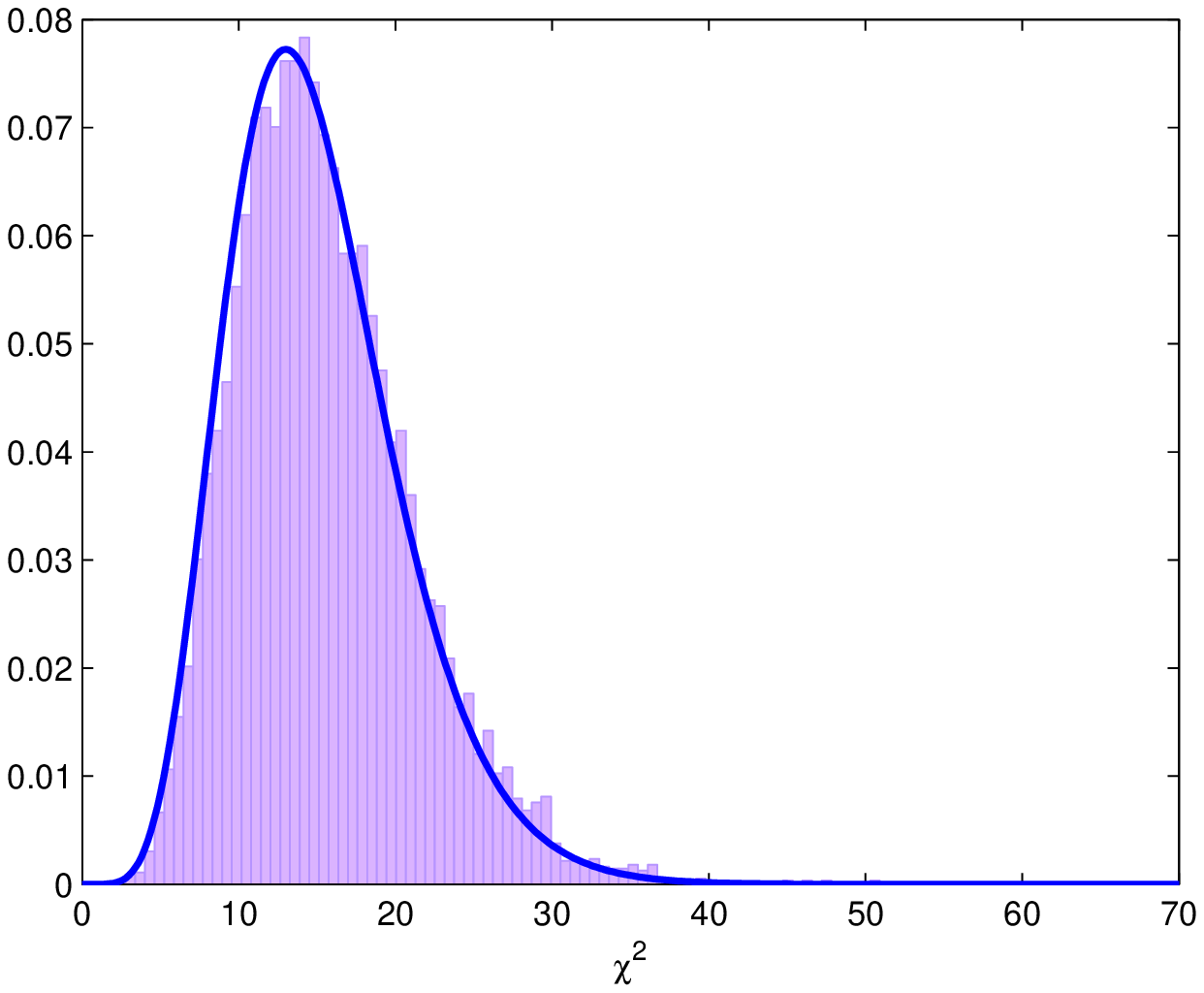} \\
\end{tabular}
\end{center}
\caption{ $\chi^2$ versus significance obtained for the 90--100~Hz band and comparison 
 of the $\chi^2$ values obtained with a $\chi^2$ distribution with $p-1$ degrees of freedom. The top figures correspond to $p = 8$ and the bottom ones to  $p = 16$.}
\label{Fig.TOTAL_QuietBand}
\end{figure}

\begin{figure}
\begin{tabular}{cc}
91.1-99.9~Hz & 101.1-101.9~Hz \\
\includegraphics[width=6.3cm]{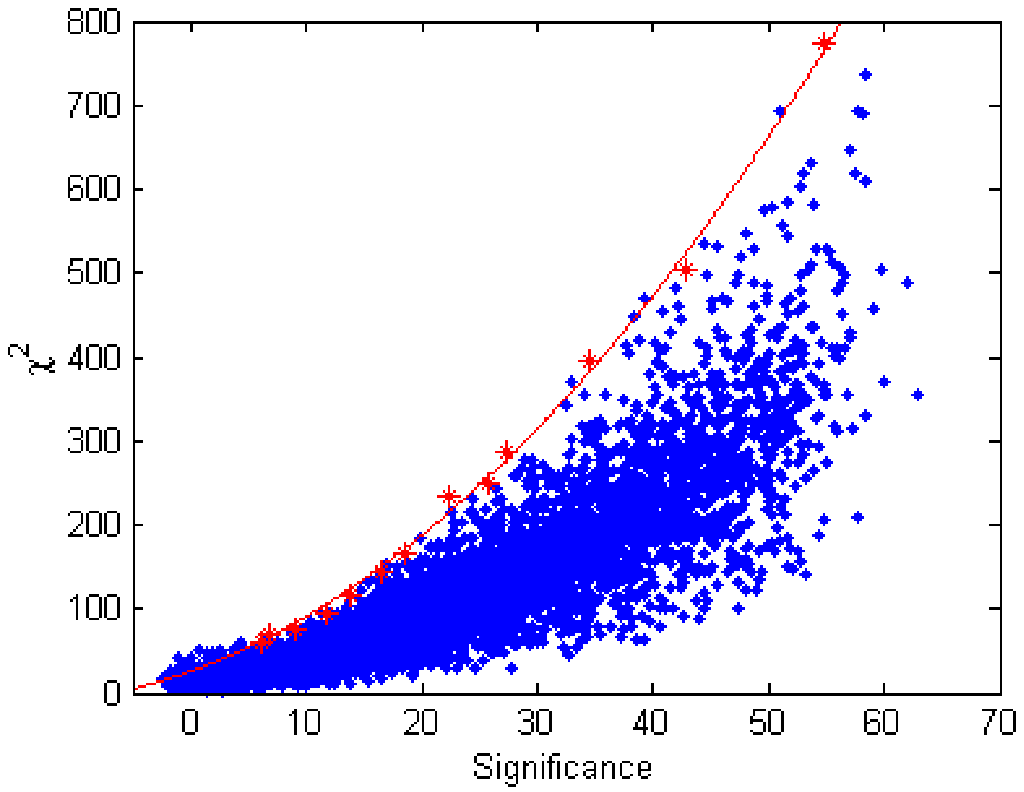} & 
 \includegraphics[width=6.3cm]{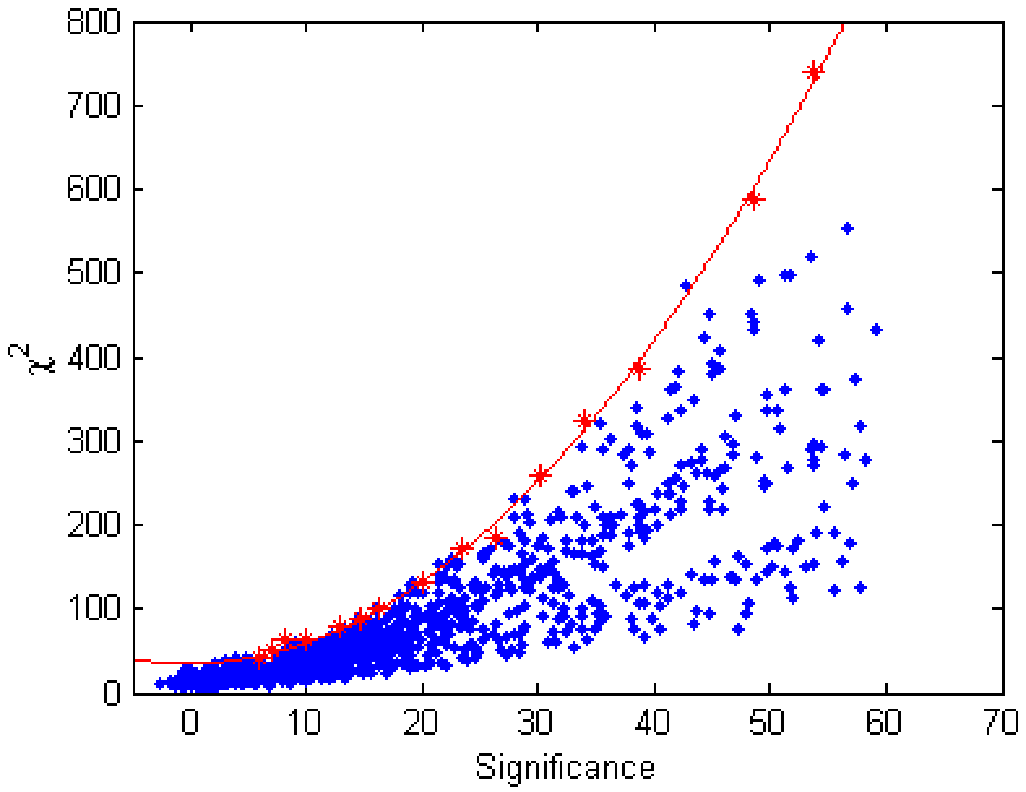} \\
 252.1-252.9~Hz & 420.1-420.9~Hz\\
\includegraphics[width=6.3cm]{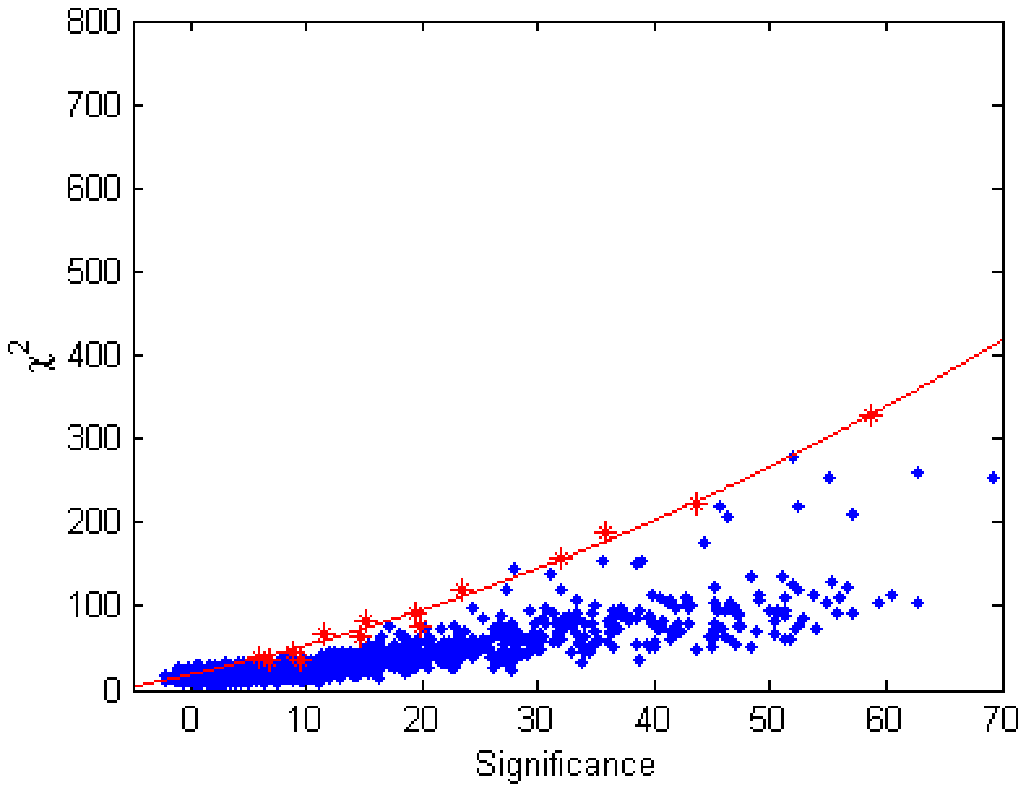} &
 \includegraphics[width=6.3cm]{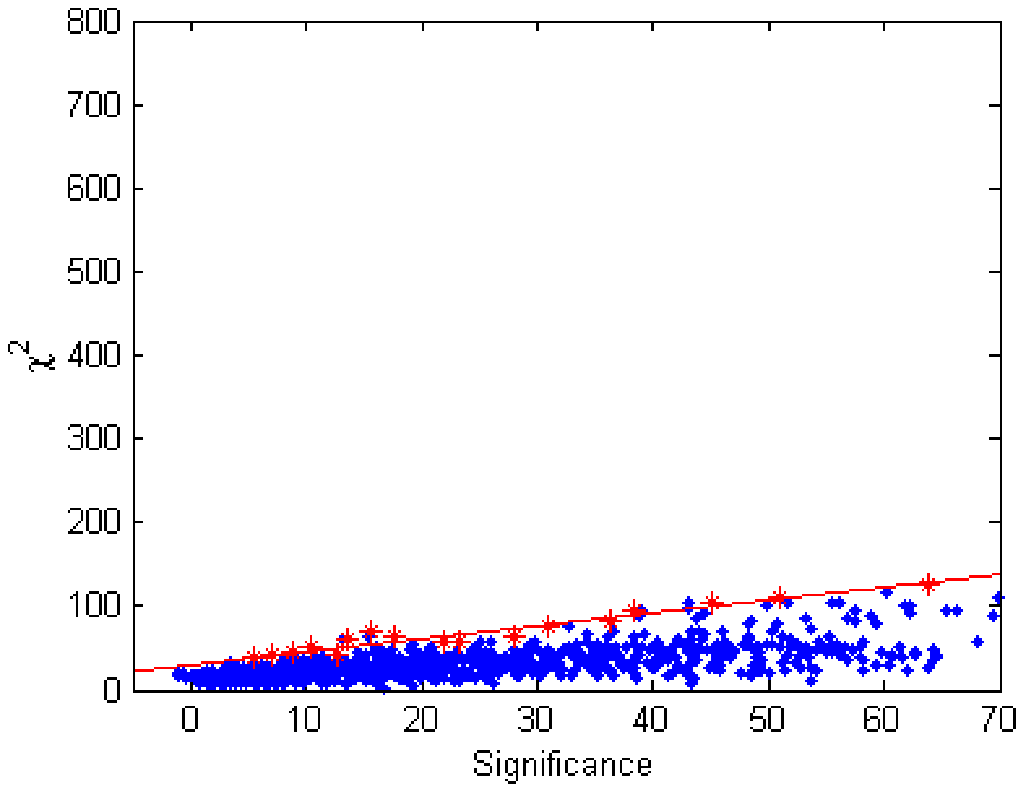} \\
\end{tabular}
\caption{ $\chi^2$ versus significance  for software injected pulsar signals. 
We represent the results for four different frequency bands of 0.8 Hz free of spectral disturbances (avoiding the 1 Hz comb) using  $p=16$, together with the best fitted quadratic curve of the envelope of the points obtained. }
\label{Fig.ChiSig_bandes}
\end{figure}

\begin{figure}
\begin{center}
\begin{tabular}{cc}
\includegraphics[width=6.3cm]{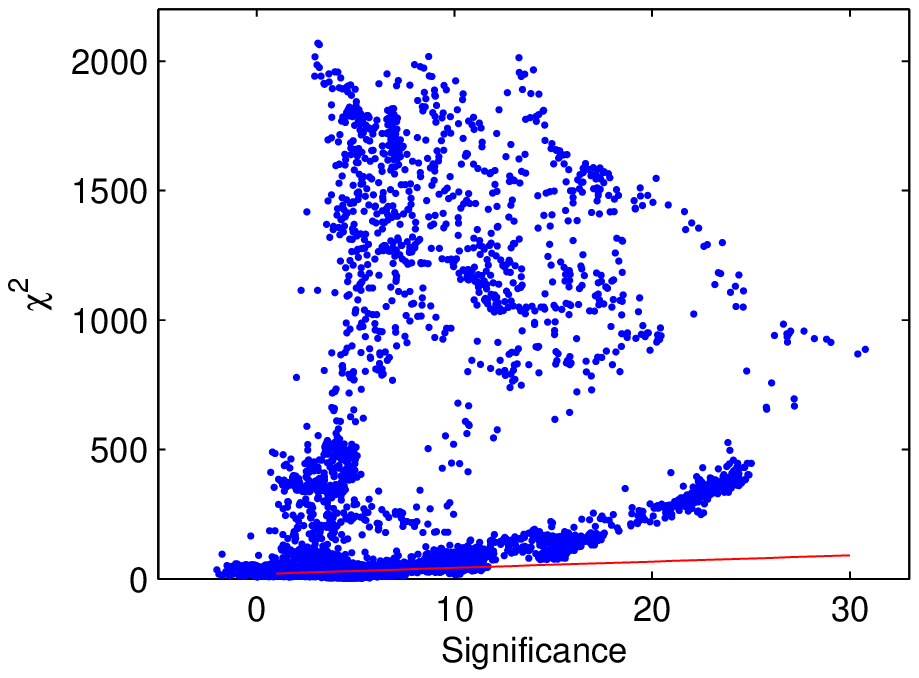} &
\includegraphics[width=6.3cm]{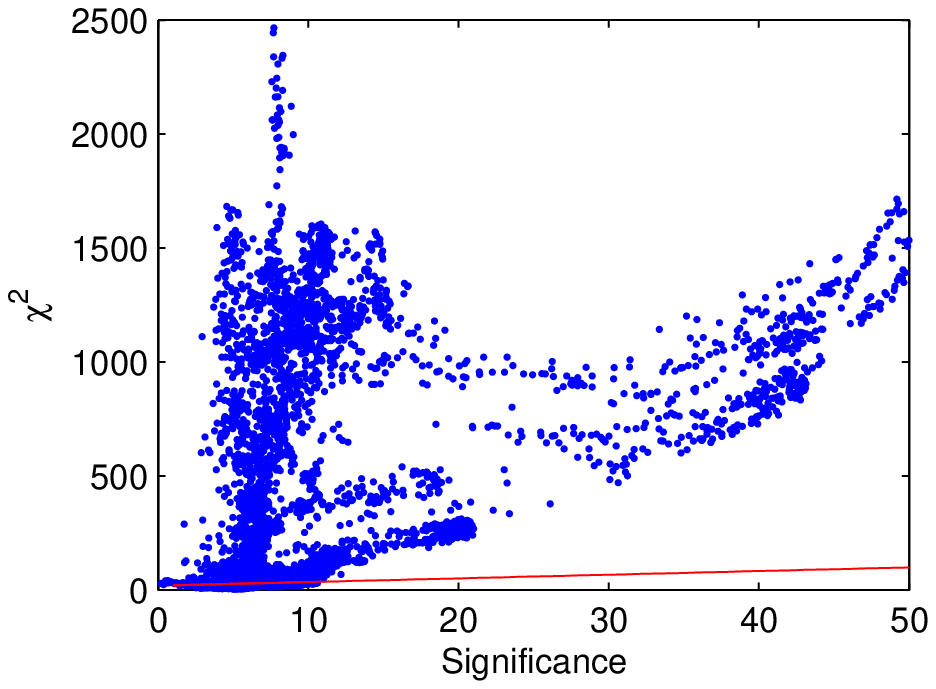} \\
\includegraphics[width=6.3cm]{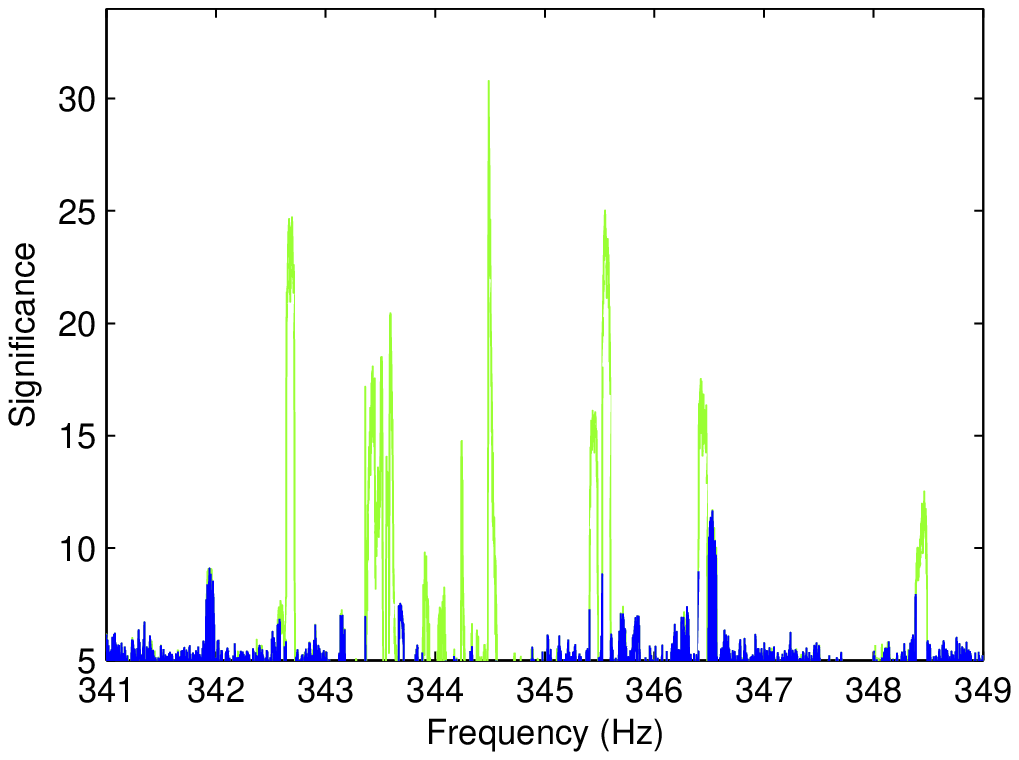} &
\includegraphics[width=6.3cm]{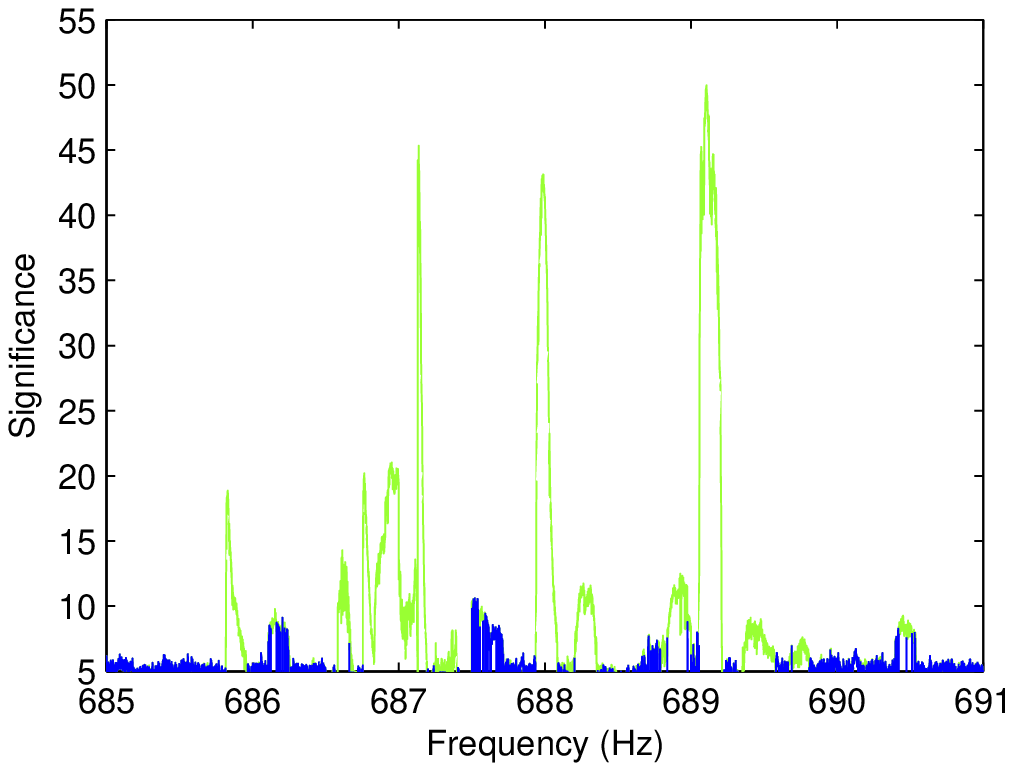} \\
\end{tabular}
\end{center}
\caption{Veto of the violin modes by the $\chi^2$ using $p = 16$. In the top panels there is the
  $\chi^2$- significance plane and  the solid line corresponds to the veto curve at an average frequency in each case. The bottom panels show the significance versus frequency before and after applying the $\chi^2$ veto. The test successfully vetoes the strongest lines related to the violin modes while it fails to veto others, e.g. the line at 342 Hz. The dots below the veto curve  with small values of significance are 
consistent with the expected distribution in the case of Gaussian noise.}
\label{Fig:veto_violin}
\end{figure}

\begin{figure}
\begin{center}
\begin{tabular}{cc}
\includegraphics[width=6.3cm]{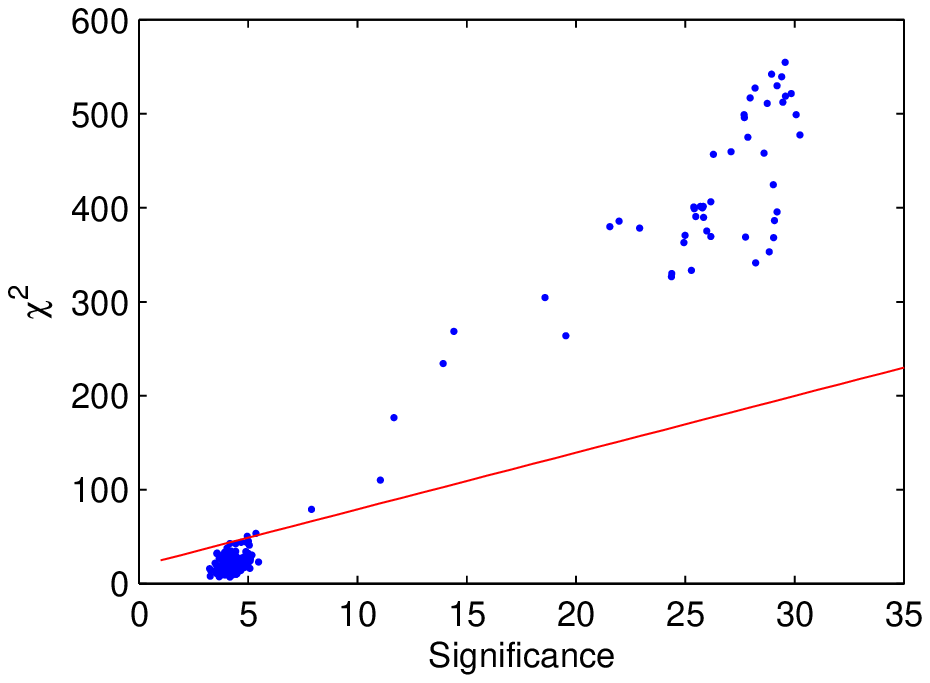} &
\includegraphics[width=6.3cm]{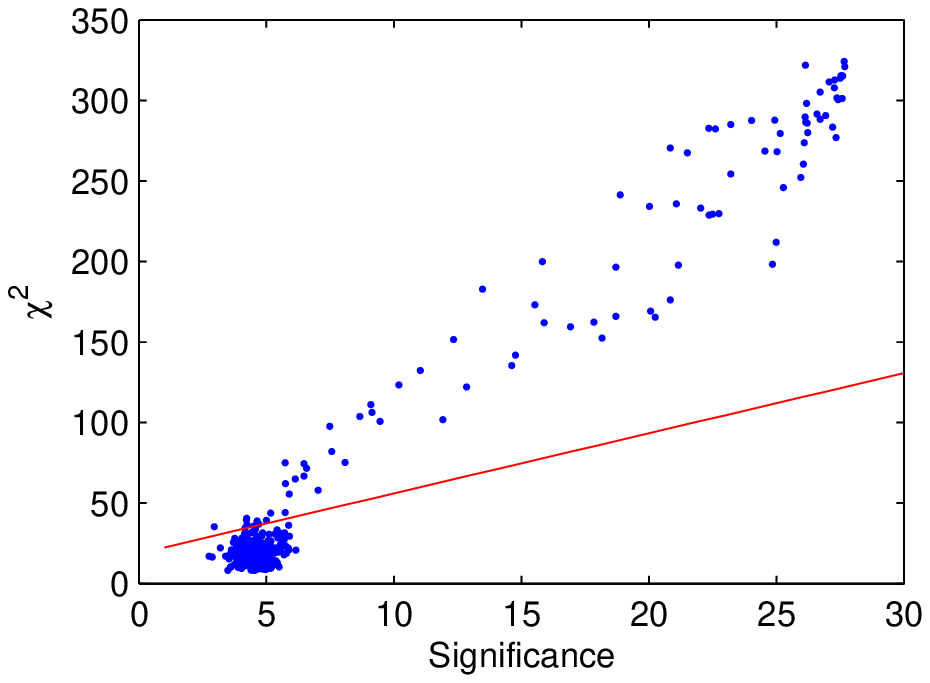} \\
\includegraphics[width=6.3cm]{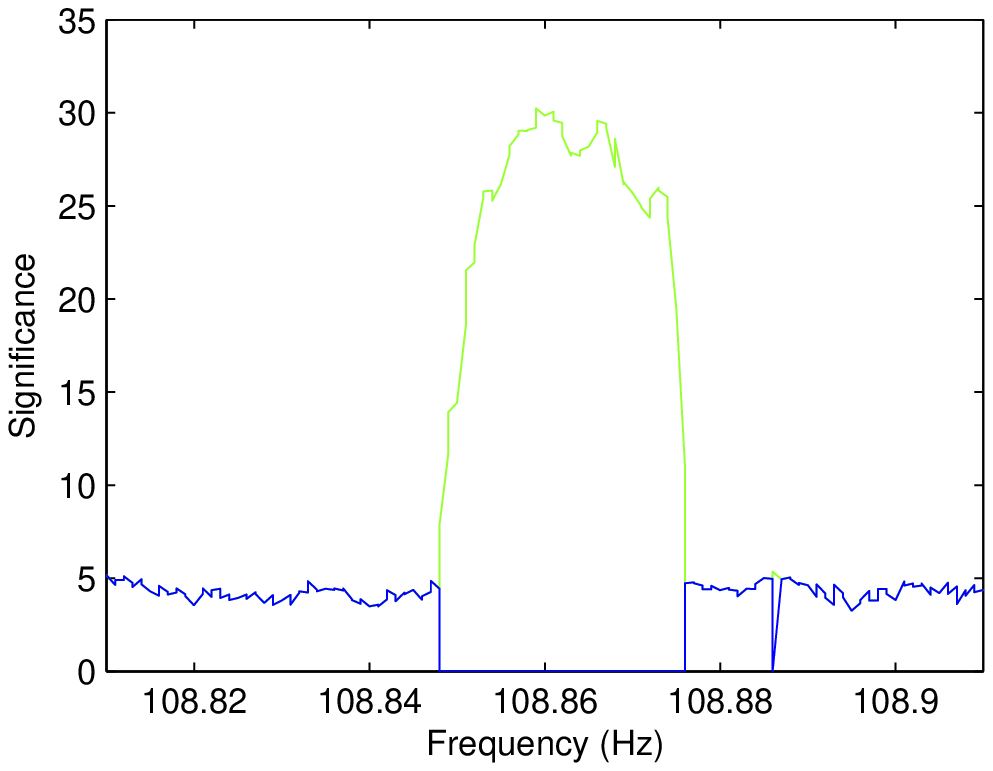} &
\includegraphics[width=6.3cm]{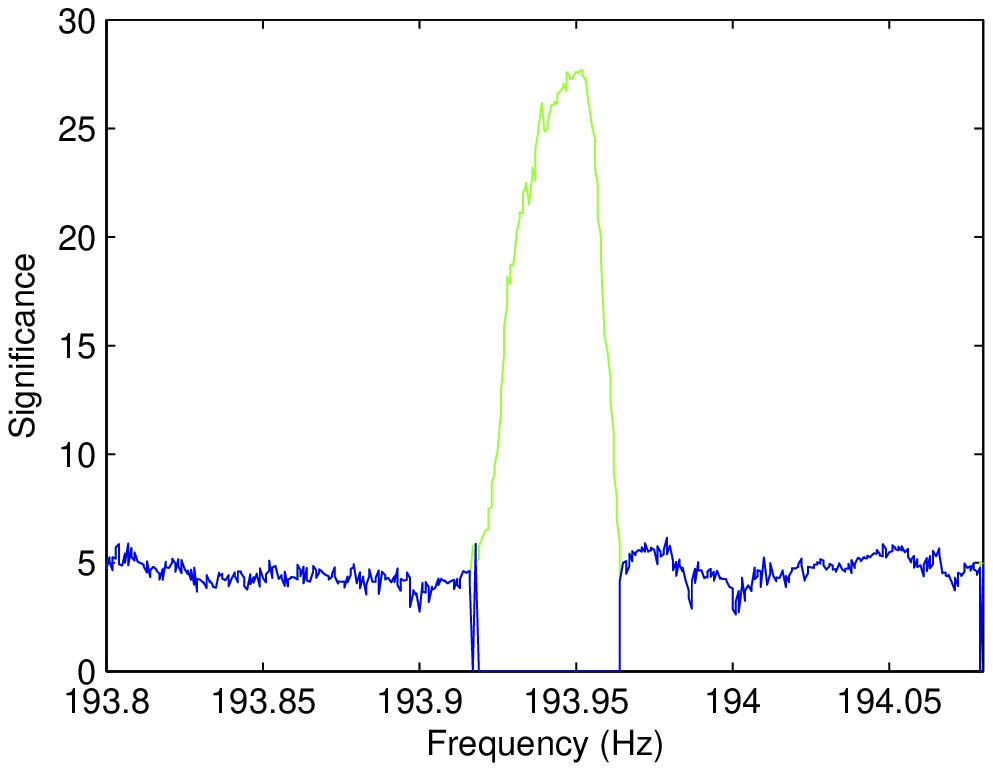} \\
\end{tabular}
\end{center}
\caption{Same as figure \ref{Fig:veto_violin} for the frequency bands that contain the hardware injected 
Pulsar 3 (left) and Pulsar 8 (right). In the top panels, the dots on top of the solid line 
correspond to the hardware injected signals that are vetoed. The dots below the curve, with low values of significance, are consistent with the expected distribution in the Gaussian noise case and correspond to the frequencies not affected by the presence of the hardware injections. The $\chi^2$ test is able to veto these signals because there were injected in the data in an intermittent way and therefore they did not behave like the signals we are looking for.}
\label{Fig:veto_pulsars}
\end{figure}

We use as playground the SFT data produced during LIGO's 29.5-day fourth science run (S4) and analyze it by means of the ``weighted Hough'' scheme, combining the data of the 
three LIGO detectors (H1, H2, L1). The SFTs were generated directly from the calibrated data stream,
 using 30-minute intervals of contiguous data for which the interferometer is
operating in what is known as science-mode, having in the end 
 1004 SFTs from H1, 1063 SFTs from H2 and  899 from L1.
An all-sky search for periodic gravitational waves in the frequency
range $50\,$--$\,1000$~Hz and with the frequency's time derivative in the 
range $-\sci{1}{-8}$--$\,0~\mathrm{Hz}~\mathrm{s}^{-1}$ using the S4 data
is reported in \cite{S4IncoherentPaper}. In that paper, three different semi-coherent 
methods of transforming and summing strain power from  SFT data
 have been used. In \cite{S4IncoherentPaper}, to identify the most interesting
  subset in parameter space, a fixed threshold of SNR or significance of 7 was
  applied for all three searches, and candidates were analyzed by means of a
  simple coincident test. The paper reported no 
   evidence of periodic gravitational radiation.

To characterize the $\chi^2$-significance
 plane in order to discriminate between instrumental noise and real signals, we first study some small 
frequency bands of the S4 data. The main purpose is to verify that in those bands 
 free of large spectral disturbances, the measured  $\chi^2$ distribution is consistent with what we would expect in the case of Gaussian noise. For this we split the data into $p=8$ or $p=16$ segments.
It is reasonable to split the data into segments with a similar relative
contribution to the total number count. Therefore, we  split the SFTs in such a way that the sum of weights into each block satisfies $\sum_{i\in I_j} \omega_{i} \approx {N}/{p}$.
In figure \ref{Fig.TOTAL_QuietBand} we show the distributions in the 90--100 Hz frequency band. 
The results agree very well with the expected theoretical distribution. 
This test was carefully done avoiding the 1 Hz comb present in 
the data.

The next step is to characterize the  $\chi^2$-significance plane in the presence of signals. 
If there was no mismatch between the signals and the templates, and if we could compute exactly the 
 $\chi^2$ value given by Eqn. (\ref{eq:chi5}) instead of (\ref{eq:chi6}),
we would obtain the same  $\chi^2$ distribution as for the Gaussian noise only case. But in a real search, templates
are placed on a grid and due to the mismatch there is a dependency of the $\chi^2$ values 
with the significance \cite{allen05}.
For this reason we select 22 frequency bands between 50 and 1000 Hz free of spectral 
disturbances and we analyze them by means of Monte-Carlo software injections. 
For each of these bands we inject at least 10000 artificial signals
of different amplitudes, frequencies, inclination angles and sky locations.
We have checked that 10000 signals was enough for our purposes. More extensive analysis were performed 
in the 91-100 Hz band, in which we injected up to 100000 signals without observing any considerable impact in the final result.
 In figure \ref{Fig.ChiSig_bandes}
we represent the results obtained for four of these different bands. 
The Monte-Carlo injections have been analyzed with no mismatch and also with a small mismatch between the signals and the templates, using the same grid separation that was employed in the S4 search 
\cite{S4IncoherentPaper}. It is worth mentioning that this grid was not based on a metric approach 
that would guarantee a maximum given mismatch at any point in parameter space, but it was uniformly spaced in frequency and  frequency derivative and used an almost isotropic grid, but frequency dependent,  in the sky. As a consequence of this choice,
together with the fact that the S4 run was shorter than a month,
the mismatch depends on the sky location and the frequency.

For each of the analyzed bands we find a veto curve.
The way we proceed is the following: we first sort the points with respect to the significance, we group them in
sets containing equal number of points (typically 50) and  we pick the one with the largest $\chi^2$ value (the highlighted stars in  Fig.  \ref{Fig.ChiSig_bandes}). With the set of selected points, we fit a curve to the upper contour of the $\chi^2$-significance planes. 
From the set of these curves we deduce empirically the best parameters of a quadratic curve 
$\chi^2= A s^2 + Bs+C$ valid for any frequency between 50 and 1000 Hz and for significance values greater than 5. For the S4 data and using $p=16$, these parameters are: 
\bea
A & =& \left \{ \begin{array}{cc}
-4.229\cdot 10^{-4}  f + 0.1274 & 50 <f< 300 \\
 0 &  300< f <1000 
\end{array} \right.  \nonumber\\
B&=& \frac{562.6}{f} + 0.7873  \nonumber\\
C &= & 18.666  \nonumber
\eea
We want to point out that this veto curve, in this case,  was only valid for significance values greater than 5. In a search, one will set a threshold on the significance  equal or higher than this value (a threshold of 7 was used for selecting triggers in the S4 search in \cite{S4IncoherentPaper}), and for each trigger one will compare if the  $\chi^2$ value is above or below the veto curve.

Using this veto curve we have analyzed the whole S4  data and here we show details of some frequency bands.
This $\chi^2$ discriminator is able to veto all the violin modes present in the data  and many other narrow instrumental lines, as it is shown in figure \ref{Fig:veto_violin}. In this figure one can see how 
the strong lines are clearly vetoed while this test failed to veto others, e.g. the line at 342 Hz.

During the S4 run ten artificial pulsar signals were hardware injected in the data in an intermittent way.
Four of these pulsars Pulsar2, Pulsar3, Pulsar8 and Pulsar9 were strong enough to be detected by the multi-interferometer Hough search (see  \cite{S4IncoherentPaper} for further details). Because  these
signals are not continuously present in the data, the $\chi^2$ test is able to veto them as it is shown  in figure \ref{Fig:veto_pulsars}. Of course, this does not happen if we analyze only the data segments when the injections took place. In those cases, the $\chi^2$-significance plane is consistent with the one obtained for software injected signals. It is worth mention that this test failed to veto all the 60 Hz line harmonics.


\section{Conclusions}
\label{sec:conc}
In this paper we have presented a  new $\chi^2$ veto adapted to the Hough transform search for continuous
gravitational wave signals  and discussed the performance of this veto using the data from LIGO fourth science run. The implementation of this veto is very simple and does not imply a considerable increase in computational cost. 
 We foresee its usage in future searches performed by the LIGO and VIRGO Scientific Collaboration. 
 The $\chi^2$ veto presented here is adapted to the Hough transform but it could be  generalized for other semi-coherent techniques such as PowerFlux or StackSlide, and also when the Hough transform starts with 
the maximum detection statistic (known as $\F$-Statistics \cite{S2FstatPaper,hough04}) rather than SFT power as the input data.

\section*{Acknowledgments}
 We would like to thank the LIGO Scientific Collaboration for many
useful discussions and for providing the data.  We also acknowledge the support of the Max-Planck
Society, the Spanish  Ministerio de Educaci\'on y Ciencia
Research Projects FPA-2007-60220, HA2007-0042, CSD207-00042 and the Govern de les Illes
Balears, Conselleria d'Economia, Hisenda i Innovaci\'o.  
We are also grateful to the Albert
Einstein Institute for hospitality where this work was initiated. The analysis were performed with the aid of the Merlin cluster of the Albert Einstein Institute.
This document has been assigned LIGO Laboratory Document No. LIGO-P080030-00-Z.




\section*{Bibliography}

\begin{thebibliography}{99}

\bibitem{GEO1}
Willke B \etal 2004
 \textit{Class.  Quant.  Grav.} \textbf{21} S417

\bibitem{GEO2} Grote H \etal 2005
  \textit{Class.  Quant.  Grav.} \textbf{22} S193

\bibitem{ligo1}
Abramovici A\etal 1992
 \textit{Science} \textbf{256} 325 

\bibitem{ligo2}
Barish B and Weiss R 1999
 \textit{Phys. Today} \textbf{52} No. 10, 44 


\bibitem{virgo97}
Caron B et al 1997
 \textit{Nucl. Phys. B-Proc. Suppl.} {\bf 54} 167 

\bibitem{S1PulsarPaper}
Abbott B \etal\ (The LIGO Scientific Collaboration) 2004
 \textit{Phys.\ Rev.}\ D {\bf 69} 082004

\bibitem{S2TDPaper}
Abbott B \etal\ (The LIGO Scientific Collaboration) 2005
 \textit{Phys.\ Rev.\ Lett.}\ {\bf 94} 181103 

\bibitem{S3S4TDPaper}
Abbott B \etal\ (The LIGO Scientific Collaboration), Kramer M
and Lyne A G  2007
 \textit{Phys.\ Rev.}\ D {\bf 76} 042001 

\bibitem{S2FstatPaper}
Abbott B\etal\ (The LIGO Scientific Collaboration) 2007
 \textit{Phys.\ Rev.}\ D {\bf 76} 082001

\bibitem{S4RadiometerPaper}
Abbott B \etal\ (The LIGO Scientific Collaboration) 2007 
 \textit{Phys.\ Rev.}\ D {\bf 76} 082003 


\bibitem{S2HoughPaper}
Abbott B\etal\ (The LIGO Scientific Collaboration) 2005
 \textit{Phys.\ Rev.}\ D \textbf{72} 102004 

\bibitem{S4IncoherentPaper}
Abbott B \etal\ (The LIGO Scientific Collaboration) 2008
 \textit{Phys.\ Rev.}\ D \textbf{77} 022001 

\bibitem{bc} Brady P R and Creighton T 2000  \textit{Phys. Rev.} D
 {\bf 61} 082001

\bibitem{pss01} Papa M A,  Schutz B F and  Sintes A M 2001, in
\textit{Gravitational waves: A challenge to theoretical astrophysics},
ICTP Lecture Notes Series, Vol. III,
edited by V. Ferrari, J.C. Miller, L. Rezzolla, Italy, p. 431

\bibitem{cgk} Cutler, C Gholami I and Krishnan B 2005
  \textit{Phys. Rev.} D \textbf{72} 042004

\bibitem{f2000} Frasca S  2000
\textit{Int. J. Mod. Phys.} D {\bf 9}  369

 \bibitem{fap} Frasca S, Astone P and  Palomba C 2005
\textit{\CQG} {\bf 22}  S1013

 \bibitem{fp} Frasca S and  Palomba C 2004
\textit{\CQG} {\bf 21}  S1645

\bibitem{EatH} \texttt{http://einstein.phys.uwm.edu/}.


\bibitem{hough04} Krishnan B, Sintes A  M,  Papa M A, Schutz B F, Frasca S and
Palomba C 2004 \textit{Phys. Rev.} D {\bf 70} 082001

\bibitem{ks07}Krishnan B and Sintes A M   2007, 
{\it Hough Search with Improved sensitivity}, LIGO Technical Document T070124, 
available in \texttt{http://admdbsrv.ligo.caltech.edu/dcc/}


\bibitem{grasp} Allen B 2000, in  \textit{GRASP: A data analysis package for
gravitational wave detection}, version 1.9.8, p.180. Available at 
\texttt{http://www.lsc-group.phys.uwm.edu/$\sim$ballen/grasp-distribution/}.

\bibitem{allen05} Allen B 2005 \textit{Phys. Rev.} D {\bf 71} 062001

\end{thebibliography}
\end{document}